\documentclass[preprint,prd,nofootinbib,tightenlines,amsmath,amssymb]{revtex4}
\usepackage{enumerate}
\usepackage{multirow}
\usepackage[utf8]{inputenc}
\usepackage{anysize}
\usepackage{textcomp}
\usepackage{epsfig}
\usepackage{graphics,color} 
\usepackage{verbatim}
\usepackage{afterpage}
\usepackage{amsmath}    
\usepackage{soul}
\usepackage{xcolor,cancel}
\DeclareUnicodeCharacter{00A0}{ }

\usepackage{blindtext}

\catcode`\@=11
\def\sla@#1#2#3#4#5{{%
 \setbox\z@\hbox{$\m@th#4#5$}%
 \setbox\tw@\hbox{$\m@th#4#1$}%
 \dimen4\wd\ifdim\wd\z@<\wd\tw@\tw@\else\z@\fi
 \dimen@\ht\tw@
 \advance\dimen@-\dp\tw@ \advance\dimen@-\ht\z@
 \advance\dimen@\dp\z@
 \divide\dimen@\tw@ \advance\dimen@-#3\ht\tw@
 \advance\dimen@-#3\dp\tw@ \dimen@ii#2\wd\z@
 \raise-\dimen@\hbox to\dimen4{%
 \hss\kern\dimen@ii\box\tw@\kern-\dimen@ii\hss}%
 \llap{\hbox to\dimen4{\hss\box\z@\hss}}}}

\def\cpto{\mathrel {\vcenter {\baselineskip 0pt \kern 0pt
    \hbox{$H_{r.f.}$} \kern 0pt \hbox{$\longrightarrow$} }}}
\def\slashed#1{%
 \expandafter\ifx\csname sla@\string#1\endcsname\relax
{\mathpalette{\sla@/00}{#1}}
\fi}
\def\declareslashed#1#2#3#4#5{%
 \expandafter\def\csname sla@\string#5\endcsname{%
#1{\mathpalette{\sla@{#2}{#3}{#4}}{#5}}}}
 \catcode`\@=12
\declareslashed{}{/}{.08}{0}{D}
 \declareslashed{}{/}{.1}{0}{A}
 \declareslashed{}{/}{0}{-.05}{k}
 \declareslashed{}{/}{.1}{0}{\partial}
 \declareslashed{}{\not}{-.6}{0}{f}

\def\lsim{\mathrel {\vcenter {\baselineskip 0pt \kern 0pt
    \hbox{$<$} \kern 0pt \hbox{$\sim$} }}}
\def\gsim{\mathrel {\vcenter {\baselineskip 0pt \kern 0pt
    \hbox{$>$} \kern 0pt \hbox{$\sim$} }}}
\def\h{{\lambda_{12}}}

\newcommand{\bea}{\begin{eqnarray}}
\newcommand{\eea}{\end{eqnarray}}

\begin{document}

\baselineskip=15pt
\preprint{}

\title{Yukawa sector for LFV in $h\to \mu\tau$ and CP violation in $h\to \tau\tau$}

\author{Alper Hayreter$^1$\footnote{Electronic address: alper.hayreter@ozyegin.edu.tr}, Xiao-Gang He$^{2,3,4}$\footnote{Electronic address: hexg@phys.ntu.edu.tw}, German Valencia$^{5}$\footnote{Electronic address: German.Valencia@monash.edu }}
\affiliation{
$^{1}$Department of Natural and Mathematical Sciences, Ozyegin University, Istanbul 34794, Turkey.\\
$^{2}$INPAC,Department of Physics and Astronomy, Shanghai Jiao Tong University, Shanghai 200240.\\
$^{3}$CTS, CASTS and Department of Physics, National Taiwan University, Taipei 10617. \\
$^{4}$National Center for Theoretical Sciences, Hsinchu 300, Taiwan\\
$^{5}$School of Physics and Astronomy, Monash University, Clayton, VIC 3800, Australia.\footnote{On leave from Department of Physics, Iowa State University, Ames, IA 50011.}
}

\date{\today}

\vskip 1cm
\begin{abstract}

The Higgs boson discovered at the LHC opened a new chapter for particle physics. Its properties need to be studied in detail to distinguish a purely standard model (SM) Higgs boson from one of many scalars in an enlarged Higgs sector. The CMS collaboration has reported a possible lepton flavor violating (LFV) signal $h\to\mu\tau$, which if confirmed, implies that the Higgs sector is larger than in the SM. New physics responsible for this type of decay may, in general, also introduce other observable effects such as charge-parity (CP) violation in $h\to \tau\tau$. We study two types of models that single out the third generation and can induce large $h \to \mu\tau$ rates with different consequences for CP violation in $h \to \tau \tau$. Predictions for the size of the CP violating couplings require knowledge of the lepton Yukawa matrices and we discuss this in the context of two different textures considering all existing constraints.

\end{abstract}

\pacs{PACS numbers: }

\maketitle

\section{Introduction}

The Higgs boson discovered at the LHC has opened a new chapter for particle physics. During the current phase of LHC running, the Higgs couplings need to be studied precisely in order to distinguish a standard model (SM) Higgs boson from a scalar forming part of an enlarged Higgs sector (beyond the SM). Yukawa interactions provide channels to probe Higgs properties in a very direct way. 
In this work we concentrate on Higgs boson decays into a charged lepton pair, such as $\tau$ and $\mu$, which can be studied at the LHC or future colliders such as FCC, ILC and CEPC. Within the SM the tau-lepton coupling to the Higgs boson is uniquely determined by its mass, the Yukawa Lagrangian being given by 
\begin{eqnarray}
{\cal L}_{Y}&=& - y_{ij} \bar{\ell}_{Li}\ e_{Rj}  \phi +\ {\rm h.c.}
\label{lyuk}
\end{eqnarray}
Here $\ell_{Li}$ is the left handed SM lepton doublet, $e_{Rj}$ the right handed  lepton singlet, $\phi$ is the scalar Higgs doublet and $i,j=1,2,3$ are generation indices. The fields $\ell_{Li}$, $e_{R_i}$ and $\phi$ transform under the SM gauge group $SU(3)_C\times SU(2)_L\times U(1)_Y$ as $(1,2,-1)$, $(1,1,-2)$ and $(1,2,1)$, respectively.

The leptons acquire a mass when electroweak symmetry is broken and the Higgs field develops a vacuum expectation value (vev) $\langle\phi\rangle=v/\sqrt{2}$, $v\approx 246$~GeV, in which case Eq.(\ref{lyuk}) becomes
\begin{eqnarray}
{\cal L}_{Y}&=&- \left(1+\frac{h}{v}\right)\frac{y_{ij} v}{\sqrt{2}}\bar{e}_{Li}\ e_{Rj}  +\ {\rm h.c.} 
\label{lyuk2}
\end{eqnarray}
The Yukawa interaction in the lepton mass eigenstate basis is obtained from Eq.(\ref{lyuk2}) with a bi-unitary transformation $S_e^\dagger (v y_{ij}/\sqrt{2}) T_e = m_i \delta_{ij}$. In this basis the Higgs-lepton couplings, given by $g_{h\ell_i\ell_j} = m_i \delta_{ij}/v$,  are proportional to the lepton masses, flavor diagonal and real. They are also CP conserving and given by the well known expression,
\begin{eqnarray} 
{\cal L}_{Y}&=&-\left(1+\frac{h}{v}\right) m_i\bar{e}_{i}\ e_{i}\;.
\label{yuksm}
\end{eqnarray}

The CMS collaboration has reported a possible lepton flavor violating (LFV) signal of the form $h\to\mu\tau$ ($\mu\tau = \mu \bar \tau + \tau \bar \mu$). If confirmed, this implies that the Higgs sector must have new flavor changing neutral current (FCNC) interactions, beyond Eq.(\ref{yuksm}). LFV Higgs decays have been discussed using the effective Lagrangian framework by a number of authors \cite{Blankenburg:2012ex,Crivellin:2015mga,Harnik:2012pb,Dorsner:2015mja,Hayreter:2016kyv,Belusca-Maito:2016axk,Herrero-Garcia:2016uab}, as well as in the context of two Higgs doublet models \cite{Botella:2014ska,Botella:2015hoa} and others \cite{Baek:2016kud}. 
When new physics introduces this type of coupling, $h\bar{e}_ie_j$, in general, it also brings in a CP violating component into the flavor diagonal Higgs-lepton couplings, which are parametrized as
\begin{eqnarray}
{g}_{he_ie_i}&=&- \frac{h}{v}{m_i}\bar{e}_i(r _{e_i} + i \tilde r_{e_i}\gamma_5){e_i}\;. \label{CPV}
\end{eqnarray}
It is well known that the simultaneous existence of scalar and pseudo-scalar couplings in Eq.(\ref{CPV}) induces a CP violating spin-spin correlation that can in principle be measured for  tau-leptons and muons, which have weak decays that analyze their polarization \cite{He:1993fd}.  In the standard treatment of this problem, one can define a density matrix $R$ for the production of polarized tau-leptons with polarization described by a unit polarization vector ${\bf n}_{\tau(\bar \tau)}$ in the $\tau(\bar \tau)$-rest frame. With the amplitude in Eq.(\ref{CPV}) the CP violating part of the density matrix is given by
\begin{eqnarray}
   R_{CP}= -N  \beta_\tau {\rm Re}(r_\tau\tilde{r}_\tau) {\vec{p}_\tau}\cdot (
     {\bf n_\tau}\times {\bf n_{\bar \tau}})\;,\label{observable}
     \label{spinspin}
\end{eqnarray}
where $N$ is a normalization constant and ${\vec{p}_\tau} $ is the three momentum direction of the tau-lepton. Beyond tree-level, $r_\tau$ and $\tilde{r}_\tau$ acquire imaginary parts and the density matrix has additional terms that we will not consider in this paper.

The existence of CP violation in the Higgs interaction may have far reaching implications for why our universe is dominated by matter over anti-matter (Baryon Asymmetric Universe - BAU). In the SM CP violation resides only in the charged current interaction of W bosons with fermions and is known to be too small to solve the BAU problem. Searches for CP violation in Higgs interactions are therefore an important topic in particle physics even if the mechanism by which new physics gives rise to LFV and CPV interactions of Higgs with fermions is not understood. 

\section{Model Independent Argument}

In a recent paper we have illustrated this discussion in a model independent manner using the Language of effective Lagrangians  \cite{Hayreter:2016kyv}. For completeness, we begin by repeating this argument before presenting in the next section an implementation within specific multi-Higgs models. 

Assuming that there are no new particles below a few hundred GeV, one can describe deviations from the SM Higgs couplings using an effective Lagrangian that respects the symmetries of the SM \cite{Buchmuller:1985jz,Grzadkowski:2010es}. At leading order, with operators of dimension six, one already finds  the following term that modifies  Eq.(\ref{lyuk}), 
\begin{eqnarray}
{\cal L}_{6}&=&- \frac{g_{ij}}{\Lambda^2}\  ( \phi^\dagger \phi) \bar{\ell}_{Li}\  e_{Rj}\ \phi +\ {\rm h.c.}
\label{dim6}
\end{eqnarray}
The matrix $g_{ij}$ is, in general, non-diagonal and complex. Expanding this Lagrangian  in combination with  Eq.(\ref{lyuk}) after electroweak symmetry breaking gives, 
\begin{eqnarray}
{\cal L}_{Y(4+6)}&=&- \left(1+\frac{h}{v}\right)\frac{y_{ij} v}{\sqrt{2}}\bar{e}_{Li}\ e_{Rj}   - \frac{v^2}{2\Lambda^2}\left(1+\frac{3h}{v}\right)\frac{g_{ij} v}{\sqrt{2}}\bar{e}_{Li}\ e_{Rj}   +\ {\rm h.c.} 
\label{fulleq}
\end{eqnarray}
As already mentioned, there is a bi-unitary transformation that diagonalizes the mass terms, which in this case are 
\begin{equation}
\left (S_e^\dagger \frac{v }{\sqrt{2}}(y+\frac{v^2}{2\Lambda^2}g ) T_e\right )_{ij} =  m_i \delta_{ij}
\label{massm}
\end{equation}
but it is evident that it no longer diagonalizes the $h\ell\ell^\prime$ couplings as 
\begin{equation}  
\left (S_e^\dagger \frac{1}{\sqrt{2}} (y+\frac{3v^2}{2\Lambda^2}g ) T_e\right )_{ij} =\frac{m_i}{v} \delta_{ij} +\frac{v^2}{\sqrt{2}\Lambda^2}(S_e^\dagger g T_e)_{ij} 
\label{fulyuk}
\end{equation}
remains an arbitrary complex matrix except for special forms of the matrix $g$. One way to check this is to use the Fritzsch ansatz  \cite{Fritzsch:1979zq,Cheng:1987rs}  for the $y_{ij}$ and treat the $g_{ij}$ as small corrections as suggested by the prefactor $v^2/\Lambda^2$. To relate the magnitudes of $h\to \tau\mu$ and $h\to \tau\tau$ one needs a specific model for the $g_{ij}$ matrix. 

If the deviations from the SM are small, as suggested by the effective Lagrangian framework, it is conventional to write Eq.(\ref{CPV}) in terms of these deviations, $r_{e_i} = 1+\epsilon_{e_i} $. Using Eq.(\ref{fulyuk}), for the tau-lepton, for example, this means,
\begin{eqnarray}
\epsilon_{e_i} &=&  \Re (S_e^\dagger g T_e)_{ii} \frac{v^2}{\sqrt{2}\Lambda^2}\frac{v}{m_i} \nonumber \\
\tilde{r}_{e_i} &=&   \Im (S_e^\dagger g T_e)_{ii} \frac{v^2}{\sqrt{2}\Lambda^2}\frac{v}{m_i}
\label{gtt}
\end{eqnarray}

The dimension six operators mentioned above are not renormalizable and it is desirable to have a UV complete theory.  In the next section we study renormalizable theoretical models in which a BSM interaction can induce large $h \to \mu\tau$ and  CP violation in   $h \to \tau \bar \tau$ decay. There are different types of models where the Higgs interaction with fermions can violate flavor and CP. We will concentrate on models that single out the third generation which we have discussed and motivated previously.  

\section{Models with FCNC and CP violation in $h \to \ell_i \bar \ell_i$}

In the SM, when diagonalizing the mass terms, the Yukawa couplings are also diagonalized, so there are no FCNC nor CP violation. However, the existence of the flavor changing couplings $h \to \ell_i \bar \ell_j$ does not necessarily imply CP violating couplings as well.  In more complicated models where the Yukawa couplings have off-diagonal entries which allow $h \to \mu \tau$ to occur, the diagonal entries may still be real implying no CP violation of the type in Eq.(\ref{CPV}). A simple way to obtain the CP violating interaction of the type in Eq.(\ref{CPV}) is to mix the scalar and pseudo-scalar components in the Higgs potential via spontaneous or explicit CP violation \cite{TDlee}. Conversely, it is also possible to have CP violation without FCNC in multi-Higgs doublet models, like the Weinberg model of spontaneous CP violation \cite{Weinberg:1976hu} which cannot induce $h \to \mu\tau$. Type-III two Higgs doublet models, on the other hand, are able to accommodate both effects. Such models have been studied  before in detail \cite{2hdmIII} and this will not be the main thrust of our paper. We will show two models motivated by treating the third generation differently from the first two generations to reduce the hierarchy problem in the Yukawa sector: the $SU(2)_l\times SU(2)_h\times U(1)_Y$ and the non-universal Left-Right $SU(2)_L\times SU(2)_R\times U(1)_{B-L}$  have the right features.  The former provides a concrete example with  flavor changing couplings $h \to \ell_i \bar \ell_j$, yet no CP violation in $h\to \tau\tau$ decay, while the latter has both LFV and CPV Higgs couplings.
 
\subsection{The $SU(2)_l\times SU(2)_h\times U(1)_Y$ model}

The $SU(2)_l\times SU(2)_h\times U(1)_Y$ model treats the first two and the third generations differently, by assuming that the 
usual $SU(2)_L$ for the first two generations is replaced by $SU(2)_l$ and for the third generation it is replaced by $SU(2)_h$. The left-handed quark doublets $Q_L$, the right-handed quark singlets $U_R$ and $D_R$, the left-handed lepton doublets $L_L$, and the right-handed charged leptons $E_R$ transform under the gauge group as
\begin{eqnarray}
&&Q^{1,2}_L: (3,2,1,1/3)\;,\;\;Q_L^3: (3,1,2,1/3)\;,\nonumber\\
&&U^{1,2,3}_R: (3,1,1,4/3)\;,\;\;D_R^{1,2,3}: (3,1,1,-2/3)\;,\nonumber\\
&&L^{1,2}_L: (1,2,1,-1)\;,\;\;L_L^3: (1,1,2,-1)\;,\;\;E^{1,2,3}_R: (1,1,1,-2)\;,
\end{eqnarray}
where the numbers in each bracket are the quantum numbers of the corresponding field under $SU(3)_C$, $SU(2)_l$, $SU(2)_h$ and $U(1)_Y$, respectively. The superscript on each field labels the generation of the fermion. The model and most of its associated phenomenology have been described in the literature before \cite{Ma:1988dn,Muller:1996dj,Chiang:2009kb,Chiang:2016qov}, here we concentrate on the scalar sector which will be responsible for the effects we want.

Symmetry breaking of $SU(2)_l\times SU(2)_h$ down to the usual $SU(2)_L$ is achieved by the vacuum expectation value (vev) $u$, of order $\cal O$(TeV), of a bi-doublet scalar $\eta: (1,2,2,0)$. The fermion masses are provided by the subsequent symmetry breaking achieved by two Higgs doublets $\Phi_1: (1,2,1,1)$ and $\Phi_2: (1,1,2,1)$ with respective vevs $v_{1,2}$ such that $v_1^2+v_2^2=v^2$. $\Phi_1$ and $\Phi_2$ only couple to the first two and the third left-handed fermions, respectively. In general this extension of the SM produces  FCNC at tree level by exchanging physical neutral Higgs scalars. It also produces FCNC due to the exchange of $Z$ and $Z^\prime$ as discussed in the literature but this effect will not concern us here. The Yukawa Lagrangian, including leptons, is given by
\begin{eqnarray}
{\cal L}_{\rm Y} &=& f^u_{ij}\bar u_{iR} \tilde \Phi^\dagger_1 Q_{jL} + g^u_{i3}\bar u_{iR} \tilde \Phi^\dagger_2 Q_{3L}+f^d_{ij}\bar d_{iR}  \Phi^\dagger_1 Q_{jL} + g^d_{i3}\bar d_{iR} \Phi^\dagger_2 Q_{3L}\nonumber\\
&+&f^e_{ij}\bar E_{iR} \Phi^\dagger_1 L_{jL} + g^e_{i3}\bar E_{iR} \Phi^\dagger_2 L_{3L} + {\rm h.c.}\;,
\end{eqnarray}
where $\tilde \Phi = i\sigma_2 \Phi$. In the above, $j$ takes values of 1 and 2, and $i$ takes values of 1, 2, and 3.
Depending on whether neutrinos are Dirac or Majorana particles,  neutrino masses can be generated by introducing right handed neutrinos $\nu_R$ to give neutrino Dirac masses. If one also allows $\nu_R$ to have a Majorana mass, then the type-I seesaw mechanism is used to give neutrino masses. 
Since $\Phi_1$ and $\Phi_2$ give masses to the first two and the third generations,  
$v_1$ should be much smaller than $v_2$ so that the hierarchy in Yukawa couplings can be reduced. 

It is convenient to work in a rotated basis for the scalar doublets $\Psi_{1,2}$ where only one Higgs boson develops a non-zero vev, with $\tan\beta = v_1/v_2$,
\begin{eqnarray}
\left ( \begin{array}{c}
\Psi_1\\
\Psi_2
\end{array}
\right ) = 
\left ( \begin{array}{rr}
c_\beta &s_\beta\\
-s_\beta&c_\beta
\end{array}
\right )
\left ( \begin{array}{c}
\Phi_1\\
\Phi_2
\end{array}
\right )\;.
\end{eqnarray}

In this basis, we have
\begin{eqnarray}
\Psi_1 = \left (\begin{array}{c}
G^\dagger\\
\frac{1}{ \sqrt{2}}(v+h +i G^0)
\end{array}
\right )\;,\;\;
\Psi_2 = \left (\begin{array}{c}
H^\dagger\\
\frac{1}{ \sqrt{2}}(H^0 +i A^0)
\end{array}
\right )\;,
\end{eqnarray}
where $G^\dagger$ and $G^0$ are the Goldstone bosons.

One can write the neutral Higgs boson couplings to charged leptons as
\begin{eqnarray}
{\cal L}_{\rm Y} = 
-\bar e_L \left ( M^e  ( 1 + \frac{h}{ v}) + (\lambda_1^e-\lambda_2^e)(H^0 - i A^0)\right ) e_R + {\rm h.c.} \nonumber
\end{eqnarray}
where 
\begin{eqnarray}
&&M^e =\frac{1}{ \sqrt{2}}( v_1\lambda^e_1 +v_2 \lambda_2^e )= \frac{v}{ \sqrt{2}}(s_\beta \lambda^e_1 + c_\beta\lambda^e_2)\;,\nonumber\\
&&
\lambda^e_1 = \left ( \begin{array}{lll}
f^{e*}_{11}&f^{e*}_{21}&f^{e*}_{31}\\
f^{e*}_{12}&f^{e*}_{22}&f^{e*}_{32}\\
0&0&0
\end{array}
\right )\;,\;\;
\lambda^e_2 = \left ( \begin{array}{lll}
0&0&0\\
0&0&0\\
g^{e*}_{13}&g^{e*}_{23}&g^{e*}_{33}
\end{array}
\right )\;.
\label{aux1}
\end{eqnarray}
Note that the structure of the model with two vevs, of which $v_1$ enters the first two diagonal elements of $\lambda^e_{1,2}$ and $v_2$ enters the third one allows one to significantly reduce the hierarchy in $f_{11}$, $f_{22}$ and $g_{33}$ as compared to the SM case by selecting $v_2>>v_1$. However, since $v_1$ contributes to both the first and the second generation masses, a (reduced) hierarchical structure 
in $f_{ij}$ and $g_{ij}$ is still needed.

Eq.(\ref{aux1}) becomes in the fermion mass eigenstate basis,
\begin{eqnarray}
{\cal L}_{\rm Y} =
-\bar e_L \left ( \hat M^e  ( 1 +\frac{h}{ v}) + \lambda^e(H^0 - i A^0)\right ) e_R  +{\rm h.c.}
\end{eqnarray}
where $M^e= S_e \hat M^e T^\dagger_e$ with $S_e$ and $T_e$ being unitary matrices and 
$\hat M^e$  the lepton mass eigenstate matrix. $\lambda^e$ is given by
\begin{eqnarray}
\lambda^e &=& S_e^\dagger (\lambda_1^e - \lambda_2^e)T_e \nonumber\\
&=& -\frac{\sqrt{2}}{ v c_\beta}\hat M^e  + (1 + \frac{s_\beta}{ c_\beta})S_e^\dagger \lambda_1^e T_e\nonumber\\
&=&\frac{\sqrt{2}}{ v s_\beta}\hat M^e - (1 + \frac{c_\beta}{ s_\beta})S_e^\dagger \lambda_2^e T_e
\;.\label{flavor}
\end{eqnarray}

The scalar $h$ is approximately the SM Higgs like particle, but it is not yet a mass eigenstate of the Higgs potential because in general, $h$ and $H$ mix with each other. On the other hand, the Higgs potential for this model is constructed with the fields $\eta$, $\Phi_{1}$, and $\Phi_2$ which are the only ones needed for symmetry breaking, and does not have mixing between the $A^0$ and the $h$ or $H^0$ states \cite{Chiang:2009kb}. The scalar mass eigenstates $h^{m1,m2}$ can then be written in terms of $h$ and $H$  with a mixing angle $\alpha$ as usual
\begin{eqnarray}
\left ( \begin{array}{c}
h\\
H
\end{array}
\right ) = 
\left ( \begin{array}{rr}
\cos\alpha & -\sin\alpha\\
\sin\alpha & \cos\alpha
\end{array}
\right )
\left ( \begin{array}{c}
h^{m_1}\\
h^{m_2}
\end{array}
\right ).
\end{eqnarray}


If we now identify $h^{m_1}$ with the 125 GeV state observed by the LHC collider,  the Yukawa coupling between  charged leptons and $h^{m_1}$ takes the form
\begin{eqnarray}
{\cal L}_{hee} = -\bar e_L \left ( \frac{\hat M^e }{ v} \cos\alpha + \lambda^e \sin\alpha \right ) e_R h^{m_1}  +{\rm  h.c.}\label{lhll} 
\end{eqnarray}

Inspecting the above equation, one sees that the $23$ and $32$ entries are non-zero in general, and thus allow $h\to \mu \tau$ to occur. Naively, one may also expect that the $33$ entry which contributes to $h \to \tau \bar \tau$ can be complex  indicating a CP violating coupling of the type in Eq.(\ref{CPV}). This is, however, not true. When diagonalizing the mass matrix above, the phase of the $33$ entry in $\lambda^e$ is automatically removed  leading to a CP conserving $h \tau \bar \tau$ coupling. To prove this, it is sufficient to show that in the mass eigenstate basis, the 33 entry of $(S^\dagger_e\lambda_2 T_e)_{33} $ is real.

From $M^e = S_e \hat M^e T_e^\dagger$, we have $(S^\dagger_e M)_{33} = (\hat M_e T^\dagger)_{33}$ which leads to
\begin{eqnarray}
c_\beta \frac{v}{ \sqrt{2}} (T_{13} g^{e*}_{13} +T_{23}g^{e*}_{23} + T_{33}g^{e*}_{33}) = m_\tau S_{33}\;.
\end{eqnarray}
At the same time, expanding $(S^\dagger_e\lambda_2 T_e)_{33} $, we obtain
\begin{eqnarray}
(S^\dagger_e\lambda_2 T_e)_{33} = (T_{13} g^{e*}_{13} + T_{23}g^{e*}_{23} + T_{33}g^{e*}_{33}) S^*_{33} = \frac{\sqrt{2}}{ v c_\beta}
m_\tau |S_{33}|^2\;.
\end{eqnarray}
Since $m_\tau$ is normalized to be real, so is $(S^\dagger_e\lambda_2 T_e)_{33} $.

To  also have CP violation in $h \to \tau\tau$ decay, one needs to modify the Yukawa structure of the model in such a way that the couplings responsible for flavor changing $ h \to e_i \bar{e}_j$ decays can not be written in the form given in Eq.(\ref{flavor}). 
This can be achieved by introducing one more Higgs doublet transforming as either $(1,2,1,1)$ or $(1,1,2,1)$. The additional fields introduce additional couplings in the Yukawa and Higgs potentials which allow the mixing of $A^0$ with $h$ and $H^0$, for example. They can also allow the resulting $h\tau\tau$ coupling to be complex from the structure of the Yukawa couplings alone. We will not pursue this avenue here, but instead  we provide a different model with the latter feature, the non-universal Left-Right model, in the next subsection.

\subsection{The Non-universal $SU(3)_C\times SU(2)_L \times SU(2)_R\times U(1)_{B-L}$ Model}\label{modworks}

The gauge group of the non-universal Left-Right model is $SU(3)_C\times SU(2)_L \times SU(2)_R\times U(1)_{B-L}$. The quantum numbers for the first two and the third generations are chosen to be different in such a way that right handed interactions are enhanced for third generation fermions and suppressed for the first two generations. This is motivated by the large top-quark mass, the possible anomalies that have been observed in $t,b$ and $\tau$ couplings \cite{Abbaneo:2001ix,Lees:2012xj,Bozek:2010xy,Abazov:2007ab}, and the stringent constraints that exist on the couplings of the lighter fermions. 
The left-handed quark doublets $Q_L$, the right-handed quark singlets $U_R$ and $D_R$, the left-handed lepton doublets $L_L$, and the right-handed charged leptons $E_R$ transform under the original gauge group as
\begin{eqnarray}
&&Q^{1,2,3}_L: (3, 2,1,1/3)\;,\;\;Q^3_R: (3,1,2,1/3)\;,\nonumber\\
&&U^{1,2}_R: (3,1,1,4/3)\;,\;\;D_R^{1,2}: (3,1,1,-2/3)\;,\nonumber\\
&&L^{1,2,3}_L: (1,2,1,-1)\;,\;\;L^3_R:(1,1,2,-1)\;,\nonumber\\
&&E^{1,2}_R: (1,1,1,-2)\;,\;\;\nu^{1,2}_R: (1, 1, 1, 0)\;.
\end{eqnarray}
The model and many aspects of its phenomenology have been discussed before in the literature \cite{He:2002ha,He:2004wr,He:2006bk,He:2009ie,He:2012zp}. Here we concentrate on the relevant scalar-lepton interactions. There are three scalar fields affecting Yukawa couplings which we list below together with their transformation properties under the gauge group,
\begin{eqnarray}
&&H_L = \left ( \begin{array}{c}
\frac{1}{ \sqrt{2}}(v_L + h_L + i A_L)\\h^-_L
\end{array}
\right ): (1,2,1,-1)\;,\nonumber\\
&&H_R = \left ( \begin{array}{c}
\frac{1}{ \sqrt{2}}(v_R + h_R + i A_R)\\h^-_R
\end{array}
\right ): (1,1,2,-1)\;,\nonumber\\
&&\phi = \left ( \begin{array}{cc}
\frac{1}{ \sqrt{2}}(v_1 + h_1 + i a_1)&h^+_2\\
h^-_1&\frac{1}{ \sqrt{2}}(v_2 + h_2 + i a_2)
\end{array}
\right ): (1,2,2,0)\;.
\end{eqnarray}
The Yukawa couplings that can be constructed with these fields are
\begin{eqnarray}
{\cal L}_{\rm Y} =&-&\left ( \bar Q_L^{1,2,3} \lambda_L^uH_L U^{1,2}_R + \bar Q^{1,2,3}_{L} \lambda^d_L\tilde H_L D^{1,2}_R+
\bar Q^{1,2,3}_L (\lambda^q \phi + \tilde \lambda^q \tilde \phi) Q^3_R\right ) + \nonumber\\
&-& \left (\bar L_L^{1,2,3} \lambda^\nu_L H_L \nu^{1,2}_R + \bar L^{1,2,3}_{L} \lambda^e_L \tilde H_L E^{1,2}_R+
\bar L^{1,2,3}_L (\lambda^l \phi + \tilde \lambda^l \tilde \phi) L^3_R\right ) + {\rm h.c}.\;,
\end{eqnarray}
where $\tilde H_L = -i\sigma_2 H^*_L$ and $\tilde \phi = \sigma_2 \phi^*\sigma_2$.

As in the previous example, the Higgs potential in this model does not allow mixing between the scalars and pseudo-scalars, therefore the 125 GeV Higgs boson will be a linear combination of $h_L$, $h_1$ and $h_2$. To find the Yukawa coupling of the 125 GeV Higgs boson to  the charged leptons, one needs to understand how $h_{L,1,2}$ couple to the charged leptons in the basis where the neutrino mass matrix has 
been diagonalized. One can write the lepton Yukawa couplings as follows
\begin{eqnarray}
{\cal L}_{Y}= - \frac{1}{ \sqrt{2}} \bar e_L[ \lambda_L^e (v_L + h_L)+ \tilde \lambda^l (v_1 + h_1) + \lambda^l(v_2 +h_2) ]e_R + {\rm h.c.}\;.
\label{lepyukLR}
\end{eqnarray}
From this we can read the charged lepton mass matrix,
\begin{eqnarray}
&&M^e =\frac{1}{ \sqrt{2}}( \lambda^e_L v_L + \tilde \lambda^l v_1+ \lambda^l v_2 ),\nonumber\\
&&
\lambda^e_L = \left ( \begin{array}{lll}
f_{11}^l&f_{12}^l&0\\
f_{21}^l&f_{22}^l&0\\
f_{31}^l&f_{32}^l&0
\end{array}
\right )\;,\;\;
\tilde \lambda^l = \left ( \begin{array}{lll}
0&0&\tilde g_{13}^l\\
0&0&\tilde g_{23}^l\\
0&0&\tilde g_{33}^l
\end{array}
\right )\;.\;\;
\lambda^l = \left ( \begin{array}{lll}
0&0&g_{13}^l\\
0&0&g_{23}^l\\
0&0&g_{33}^l
\end{array}
\right )\;.
\end{eqnarray}

It is convenient to work in a basis where only one Higgs has non-zero vev  $v = (v_L^2 + v^2_1 +v^2_2)^{1/2}$. To do so we define
\begin{eqnarray}
\left (\begin{array}{c}
h_L\\ h_1\\ h_2
\end{array} \right ) 
= \left (\begin{array}{ccc}
v_L/v&0&v'/v\\
v_1/v&v_2/v'&-v_Lv_1/v'v\\
v_2/v&-v_1/v'&-v_Lv_2/v'v
\end{array} \right )
\left (\begin{array}{c}
\tilde h \\ H_1\\ H_2
\end{array} \right )\;.
\end{eqnarray}
where $v' = (v_1^2+v_2^2)^{1/2}$.

Assuming $S_e$ and $T_e$ diagonalize the charged lepton mass matrix, $S^\dagger M^e T = \hat M^e$, we have Eq.(\ref{lepyukLR}) in the charged lepton mass eigenstate basis as
\begin{eqnarray}
{\cal L}_{\rm Y_e} &=&- \bar e_L \left ( \hat M^e  ( 1 + \frac{\tilde h}{ v}) + \lambda^e_1 H_1+ \lambda^e_2 H_2 \right ) e_R + {\rm h.c.}\;
\end{eqnarray}
where the matrices parametrizing the lepton Yukawa couplings are now
\begin{equation}
\lambda^e_1=\frac{S_e^\dagger (\tilde \lambda^l v_2-\lambda^l v_1) T_e}{\sqrt{2}v'},\;\; \lambda^e_2 = \frac{S_e^\dagger (\lambda^e_L v' - \tilde \lambda^l \frac{v_1v_L}{ v'} - \lambda^l \frac{v_2v_L}{ v'})   T_e}{\sqrt{2} v}.
\end{equation}

The Higgs mass eigenstates can now be written as linear combinations of  $\tilde h,\;H_1,\; H_2$ as $h^{m_i} = V^{ih} \tilde h + V^{i1}H_1+V^{i2}H_2$ in terms of  an orthogonal matrix $V^{ij}$. Further identifying the lightest mass eigenstate $h^{m_1} =h$ with  the 125 GeV Higgs boson, we have
\begin{eqnarray}
{\cal L}_{he_ie_j} = \left( \frac{\hat M^e}{ v}V^{1h} + \lambda^e_1V^{11}+ \lambda^e_2 V^{12}  \right)_{ij}\, \bar e_{L_i}e_{R_j}h\;.
\label{gij}
\end{eqnarray}
In terms of the generic parameters defined in Eq.(\ref{fulyuk}), we have
\begin{equation}
\frac{v^2}{\sqrt{2}\Lambda^2}(S_e^\dagger g T_e)_{ij}= \left(\frac{\hat M^e}{ v}(V^{1h} -1)+ \lambda^e_1V^{11}+ \lambda^e_2 V^{12}  \right)_{ij},
\label{yukmod}
\end{equation}
and the normalized tau couplings $\epsilon_\tau$ and $\tilde r_\tau$ defined in Eq.(\ref{CPV}) are then\begin{eqnarray}
&&\epsilon_\tau = V^{1h}_{33}-1 +\Re\left ((\lambda^e_1)_{33}V^{11} + (\lambda^e_2)_{33} V^{12}\right )\frac{v }{ m_\tau}\;,\nonumber\\
&&\tilde r_\tau = \Im\left ((\lambda^e_1)_{33}V^{11} + (\lambda^e_2)_{33} V^{12}\right ) \frac{v }{ m_\tau}\;.
\label{defsinlr}
\end{eqnarray}

Eq.(\ref{gij}) is similar to Eq.(\ref{lhll}). However, this time there are two terms in ${\cal L}_{he_ie_j}$ which are non-diagonal. This difference is sufficient to reach opposite conclusions to the previous model:  in the mass eigenstate basis $h$ can decay to $\mu\tau$ and at the same time the Yukawa coupling for $h$ to $\tau \bar \tau$ can be complex leading to CP violating coupling of the type in Eq.(\ref{CPV}). This model naturally has non-zero values for $r_\tau$ and $\tilde r_\tau$ simultaneously. The generic scale of new physics $\Lambda$ in this model is related to the masses of the heavier scalars.

\section{Existing Constraints  and CP violation in $h\to \tau \tau$}

Inspecting Eq.(\ref{fulyuk}), one might think that the hierarchical structure of the lepton mass matrix is already  encoded in the first term so that the flavor structure of the dimension six term, $(S_e^\dagger g T_e)_{ij}$, or Eq.(\ref{yukmod}) within the model of Section(\ref{modworks}), could be democratic. Furthermore, within the models we are discussing we can choose appropriate values for $v_{L, 1,2}$ to reduce the hierarchical structure of the Yukawa couplings making  democratic $\lambda^e_{1,2}$ matrices plausible. We would write in this case,
\begin{equation}
(S_e^\dagger g T_e)_{ij}\sim \lambda^e_{1,2} \sim 
\left(
\begin{array}{ccc}
1  & 1  &  1 \\
 1 & 1  &  1 \\
  1&   1&   1
\end{array}
\right)
\;. \label{option1}
\end{equation}
and the diagonal elements, in the notation of Eq.(\ref{gtt}) would satisfy
\begin{equation}
\frac{ (\epsilon_i + i\tilde{r}_i)}{(\epsilon_j + i\tilde{r}_j)}\sim \frac{m_j}{m_i}.
\end{equation}

However, this may not be the case. For example, as mentioned before, one still needs to split the first and second generations and hierarchical $\lambda^e_{1,2}$ matrices may still be needed. In this case we could write following Refs.~\cite{Fritzsch:1979zq,Harnik:2012pb},
\begin{equation}
(S_e^\dagger g T_e)_{ij}\sim \lambda^e_{1,2} \sim 
\left(
\begin{array}{ccc}
 m_e & \sqrt{m_e m_\mu}  &  \sqrt{m_e m_\tau} \\
 \sqrt{m_e m_\mu} & m_\mu  &  \sqrt{m_\mu m_\tau} \\
  \sqrt{m_e m_\tau}&   \sqrt{m_\mu m_\tau}&   m_\tau
\end{array}
\right)\;.\label{option2}
\end{equation}
and this time the diagonal elements, in the notation of Eq.(\ref{gtt}) would satisfy
\begin{equation}
\frac{ (\epsilon_i + i\tilde{r}_i)}{(\epsilon_j + i\tilde{r}_j)}\sim 1.
\end{equation}

We will consider the above two cases as benchmarks  for discussion in the remaining of the paper. 

\subsection{Constraints on Yukawa couplings and $h \to \mu \tau$}

To explain the CMS data for $h \to \mu \tau$, it is necessary to have non-zero $g_{\mu\tau}$ and $g_{\tau\mu}$.
We have
\begin{eqnarray}
{\cal L}_{\mu\tau}&=&- (g_{h\mu\tau}\bar \mu_L \tau_R + g_{h \tau \mu} \bar \tau_L \mu_R)h - (g^*_{h\mu\tau}\bar \tau_R \mu_L + g^*_{h \tau \mu} \bar \mu_R \tau_L)h\nonumber\\
&=& -\left ( \frac{g_{h\mu\tau} + g_{h\tau\mu}^*}{ 2}\bar \mu \tau +\frac{g_{h\mu\tau}-g^*_{h\tau\mu}}{ 2} \bar \mu \gamma_5 \tau\right)h \nonumber\\
&&-\left ( \frac{g_{h\tau\mu} + g_{h\mu\tau}^*}{ 2}\bar \tau \mu +\frac{g_{h\mu\tau}-g_{h\tau\mu}^*}{ 2} \bar \mu \gamma_5 \tau\right)h\;.
\end{eqnarray}
Including loop effects, $g_{h\mu\tau}$ may have non-zero absorptive part which leads a rate difference between $h \to \bar \mu \tau$ and $h\to \bar \tau \mu$. However, if the absorptive parts are small, the rate for $h \to \bar \mu \tau$ and $h\to \bar \tau \mu$ will be approximately equal.

These couplings have been studied in connection with the CMS report \cite{Khachatryan:2015kon}
\begin{equation}
 B(h\to \mu\tau)=(0.84_{-0.37}^{+0.39})\% .
\end{equation}
When the absorptive parts in $g_{hij}$ are neglected, one obtains 
\begin{equation}
\sqrt{g_{h\tau\mu}^2+g_{h\mu\tau}^2}< 3.6\times 10^{-3}.  
\label{cmsb}
\end{equation}
The two benchmark flavor structures in Eqs.(\ref{option1}) and (\ref{option2}) thus imply
\begin{itemize}
\item democratic 
\begin{eqnarray}
\sqrt{|\epsilon_\tau|^2+|\tilde{r}_\tau|^2} &\leq& \frac{1}{\sqrt{2}}3.6 \times 10^{-3}\frac{v}{m_\tau}
\end{eqnarray}
\item hierarchical
\begin{eqnarray}
\sqrt{|\epsilon_\tau|^2+|\tilde{r}_\tau|^2} &\leq&  \frac{1}{\sqrt{2}}3.6 \times 10^{-3}\sqrt{\frac{m_\tau}{m_\mu}}\frac{v}{m_\tau}
\end{eqnarray}
\end{itemize}

We can also use the measured rates $h\to \tau\tau$ and $h\to \mu\mu$ from the ATLAS-CMS combination \cite{atlas-cms}  assuming that there is no new physics. These are,
\begin{eqnarray}
|\kappa_i|^2 &\equiv& \frac{\Gamma(h\to\ell_i\ell_i)}{\Gamma(h\to\ell_i\ell_i)_{SM}}\nonumber \\
\kappa_\tau &=& \sqrt{(1+\epsilon_\tau)^2+\tilde{r}_\tau^2}=0.90^{+0.14}_{-0.13} \nonumber \\
\kappa_\mu&=&\sqrt{(1+\epsilon_\mu)^2+\tilde{r}_\mu^2}=0.2^{+1.2}_{-0.2}
\end{eqnarray}
as well as the constraint on $h\to e^+e^-$ at 95\%c.l. \cite{Altmannshofer:2015qra}
\begin{equation}
\kappa_e=\sqrt{(1+\epsilon_e)^2+\tilde{r}_e^2} \leq 611
\end{equation}

The two flavor structure benchmarks then imply at 95\% c.l., using the notation for mass ratios $x_\mu=m_\tau/m_\mu$ and $x_e=m_\tau/m_e$
\begin{itemize}
\item democratic
\begin{eqnarray}
0.645 \leq \sqrt{|1+\epsilon_\tau|^2+|\tilde{r}_\tau|^2} &\leq& 1.174 \nonumber \\
\sqrt{|1+x_\mu\epsilon_\tau|^2+|x_\mu\tilde{r}_\tau|^2} &\leq& 2.55 \nonumber \\
\sqrt{|1+x_e\epsilon_\tau|^2+|x_e\tilde{r}_\tau|^2} &\leq& 611 
\end{eqnarray}
\item hierarchical
\begin{eqnarray}
0.645 \leq \sqrt{|1+\epsilon_\tau|^2+|\tilde{r}_\tau|^2} &\leq& 1.174 \nonumber \\
\sqrt{|1+\epsilon_\tau|^2+|\tilde{r}_\tau|^2} &\leq& 2.55 \nonumber \\
\sqrt{|1+\epsilon_\tau|^2+|\tilde{r}_\tau|^2} &\leq& 611 
\end{eqnarray}
\end{itemize}

For the LFV violating coupling, there is also a constraint from $\tau \to \mu \gamma$. From $Br(\tau \to \mu \gamma)_{exp} < 4.4\times 10^{-8}$ \cite{Agashe:2014kda}, the allowed range 
encompasses  $2.0\times 10^{-3}< \sqrt{g_{h\tau\mu}^2+g_{h\mu\tau}^2}< 3.3\times 10^{-3}$ \cite{jusak} and yields  a 95\% upper bound that is four times larger (weaker constraint) than the CMS result quoted above.

With some relatively weak constraints on $r_\tau$ and $\tilde r_\tau$, one may wonder whether a large $\tau$ edm $d_\tau$ can be generated. We have checked this possibility and found that since the contribution to $d_\tau$ from Eq.(\ref{CPV}) is proportional to $m_\tau^3(r_\tau\tilde r_\tau)/16\pi^2 v^2m_h^2$, the current upper limit $d_\tau < 10^{-17}$ e.cm does not constrain $r_\tau\tilde r_\tau$ significantly.

\begin{figure}[h]
\includegraphics[scale=0.8]{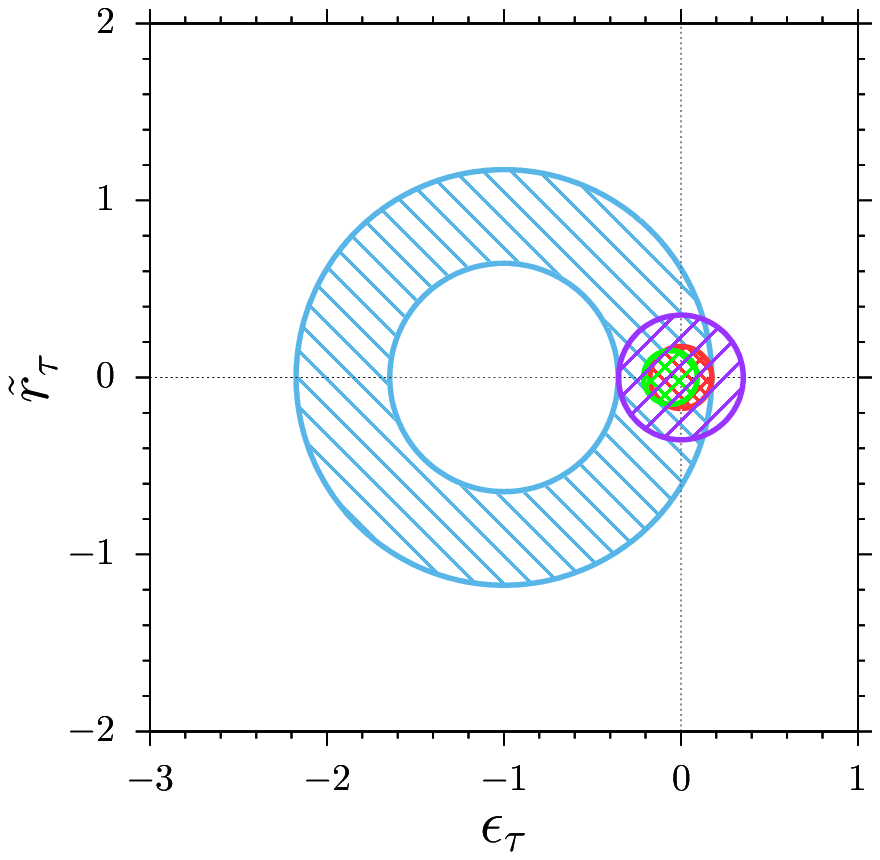} \hspace{0.5cm}
\includegraphics[scale=0.8]{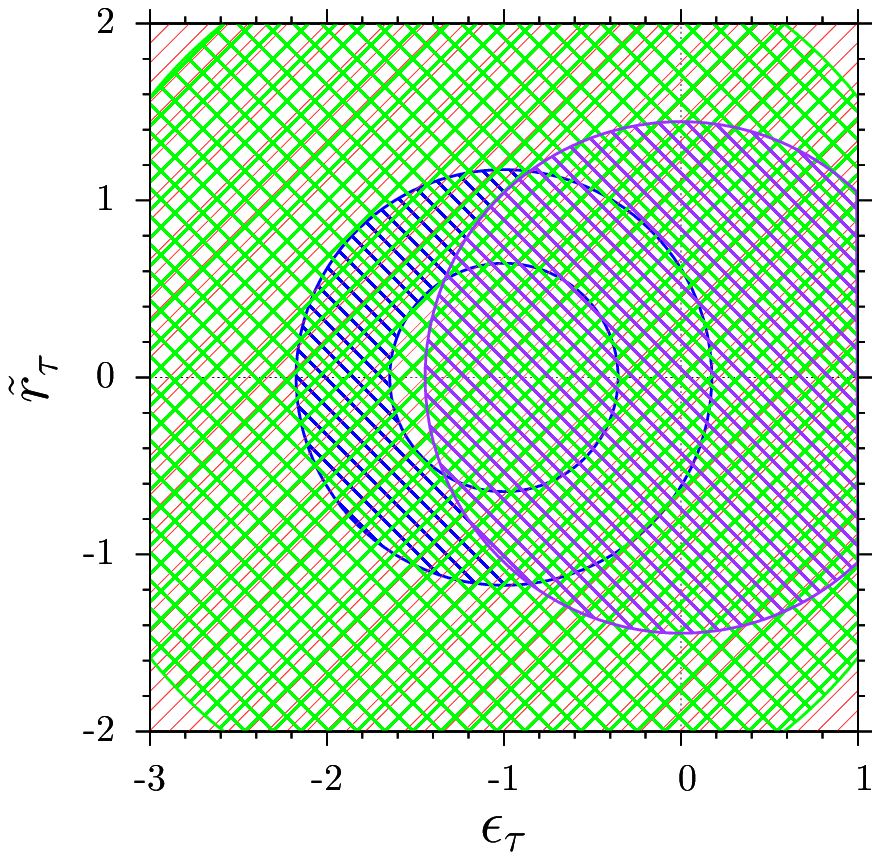} 
\caption{Region of parameter space allowed by the constraints discussed in the text at the 95\% c.l.  The blue region is from the $h\to \tau\tau$ rate, the green region from the $h\to\mu\mu$ limit, the red region is from the $h\to e e$ limit and the purple region is from the CMS $h\to\tau\mu$ upper bound.  The left panel corresponds to the democratic flavor scenario and the right panel to the hierarchical one.}
\label{fig}
\end{figure}

Our numerical constraints are summarized in Figure~\ref{fig}. The panel on the left corresponds to the democratic flavor scenario. We see in this case that the most restrictive bounds arise from the limits on $h\to\mu\mu$ and $h\to ee$. This is due to the much smaller SM Yukawa couplings for electrons and muons relative to tau-leptons, which significantly enhance the effects of democratic absolute deviations from the SM in the relative couplings $\kappa_i$ probed by experiment. We see in this case that the maximum allowed value of the ratio that quantifies CP violation is 
\begin{equation}
\left| \frac{r_\tau \tilde{r}_\tau  }{r_\tau^2 + \tilde{r}_\tau^2  }\right|\leq 0.15.
\end{equation}
The panel on right corresponds to the hierarchical flavor scenario. In this case the most restrictive constraints are those from $h\to \tau\tau$ and from the global fit for $\epsilon_\tau$. This case still allows the CP violating ratio to take its maximum possible value
\begin{equation}
\left| \frac{r_\tau \tilde{r}_\tau  }{r_\tau^2 + \tilde{r}_\tau^2  }\right|\leq 0.5.
\end{equation}
Constraints that can be placed on these couplings in future colliders have been recently investigated in Ref.~\cite{Banerjee:2016foh}, and in the next section we compare some possible CP-odd asymmetries for that purpose.

\subsection{CP violation in $h\to \tau \bar \tau$}

As discussed above, the couplings $r$ and $\tilde r$ in Eq.(\ref{CPV}) give rise to a CP violating spin-spin correlation as in Eq.(\ref{spinspin}). The polarizations of $\tau$ and $\bar \tau$ can be extracted in principle by studying the angular distributions of their decay. In this section we will study the relative sensitivity of the different tau-lepton decay modes to CP violation at a more theoretical level by comparing the T-odd correlations for each case. Experimental study of these correlations requires the reconstruction of the Higgs rest frame, which in the di-tau mode, is not possible at LHC. They are thus better suited for study at an $e^+e^-$ collider. For example, an ILC or CPEC running at 250~GeV would produce the Higgs through the $e^+e^-\to Zh$ reaction and modes that reconstruct the $Z$ completely (such as the di-muon mode) will allow full reconstruction of the Higgs rest frame \cite{Han:2000mi,Han:2013kya,Gomez-Ceballos:2013zzn}. A full phenomenological analysis of how these asymmetries could be measured is beyond the scope of this paper. We present them here simply as an illustration of the type of effects that can be expected.

The simplest mode to consider is the two body decay already discussed in Refs.~\cite{He:1993fd,Hayreter:2016kyv}
\begin{equation}
\tau^- \rightarrow \pi^-\nu_{\tau},\ \ \tau^+ \rightarrow \pi^+\bar \nu_{\tau}\;.
\end{equation}
Denoting by $\vec{p}_{\pi^\pm}$ the three- momenta of the pions in the Higgs rest frame, Eq.(\ref{spinspin}) generates the T-odd correlation
\begin{eqnarray}
{\cal O}_\pi = \vec{p}_\tau\cdot(\vec{p}_{\pi^+} \times \vec{ p}_{\pi^-}).
\end{eqnarray}
This can be measured, for example, by the integrated counting asymmetry
\begin{eqnarray}
A_\pi =\frac{N({\cal O}_\pi >0)-N({\cal O}_\pi <0) }{N({\cal O}_\pi >0)+N({\cal O}_\pi<0 )}=
\frac{ \pi}{ 4} \beta_\tau \frac{ (r_\tau \tilde{r}_\tau )}{\beta_\tau^2  r_\tau^2 + \tilde{r}_\tau^2  },
\label{piasym}
\end{eqnarray}
as has been known for a long time \cite{He:1993fd}.

The asymmetry for the leptonic three body decay $\tau^\pm \to \ell^\pm \nu\bar{\nu}$  can also be calculated analytically. In this case it is simplest to directly construct the triple product correlation between final particle momenta using the methods of Ref.~\cite{Antipin:2008zx} to compute the relevant density matrices and obtain the Lorentz invariant form of the CP violating matrix element squared,
\begin{eqnarray}
\left|{\cal M}_{\slashed{CP}}\right|&=&-\frac{32\pi^2r_\tau\tilde{r}_\tau}{\Gamma_\tau^2}\left(4\sqrt{2}G_F\right)^2
\delta(p_{\tau^+}^2-m_\tau^2)\delta(p_{\tau^-}^2-m_\tau^2){\cal O}\nonumber \\
{\cal O}&=&\epsilon^{\mu\nu\alpha\beta}p^\mu_{\tau^-}p^\nu_{\tau^+}p^\alpha_{\nu_\ell}p^\beta_{\bar\nu_\ell}
\end{eqnarray}
The delta functions reveal that we have used the narrow-width approximation for the denominator of the tau-lepton propagators, but we have kept all spin correlations, and $\cal O$ is the Lorentz invariant form of the raw CP violating correlation that occurs in this decay. In the same way we can calculate the total decay width for this channel,  with $\beta_\tau=\sqrt{1-4m_\tau^2/m_h^2}$ we find
\begin{eqnarray}
\Gamma= \frac{\beta_\tau}{8\pi m_H}m_\tau^2\left(\frac{m_H^2}{v^2}\right)\left(\beta_\tau^2|r_\tau|^2+|\tilde{r}_\tau|^2\right)Br(\tau\to\mu+\nu's)^2.
\end{eqnarray}
To measure the CP odd correlation we would use an integrated counting asymmetry
\begin{equation}
A=\frac{N_{\rm ev}({\cal O}>0)-N_{\rm ev}({\cal O}<0)}{N_{\rm ev}({\cal O}>0)+N_{\rm ev}({\cal O}<0)}.
\end{equation}
In  the limit $m_\tau<<m_H$, $\beta_\tau\to 1$ and $m_\ell << m_\tau$ it is possible to compute this analytically by
integrating over the six body phase space as sketched in Ref.~\cite{Valencia:2005cx}, resulting in
\begin{eqnarray}
A&=&- \frac{\pi}{4}\frac{r_\tau\ \tilde{r}_\tau}{|r_\tau|^2+|\tilde{r}_\tau|^2}.
\label{nuasym}
\end{eqnarray}
Of course this is just the raw asymmetry as the neutrino momenta cannot be measured. It represents the largest possible asymmetry in this mode as there are dilution factors when the triple product is projected onto observable momenta. This part of the calculation is better done numerically and to this aim we  implemented the Lagrangian of  Eq.(\ref{CPV}) in FEYNRULES \cite{Christensen:2008py,Degrande:2011ua} to generate the Universal Feynrules Output (UFO) file, then feeding this UFO file into MG5\_aMC@NLO \cite{Alwall:2014hca} in combination with TAUDECAY \cite{Hagiwara:2012vz} package which performs the hadronic decays of the tau-lepton. A suitable T-odd correlation for the leptonic decay mode is
\begin{eqnarray}
{\cal O}_\ell = \vec{p}_\tau\cdot(\vec{p}_{\ell^+} \times \vec{ p}_{\ell^-}),
\end{eqnarray}
where now $\vec{p}_{\ell^\pm}$ denotes the three-momenta of the charged lepton in the Higgs rest frame, and can be measured with the integrated counting asymmetry 
\begin{eqnarray}
A_{\ell}&=& \frac{\pi}{36}\frac{r_\tau\ \tilde{r}_\tau}{|r_\tau|^2+|\tilde{r}_\tau|^2}.
\label{lepasym}
\end{eqnarray}

For more than one pion in the decay of $\tau$'s, we have carried out a similar analysis with results sumarized in Tables~\ref{tabl}~and~\ref{tabh}. For all cases in the Tables, we simulated the Higgs boson decay in its rest frame with 200000 events with no kinematic cuts for a sufficient number of values $r_\tau$, $\tilde{r}_\tau$ to obtain a good fit to the asymmetry. For modes with more than one pion we measured different T-odd correlations using the different pion momenta available, but in all cases studied found that the largest sensitivity was obtained by using a `tau-jet' momenta defined as the sum of all the pion momenta in the corresponding decay.

Table~\ref{tabl} shows the semi-leptonic modes $h\to \tau^+\tau^- \to \tau_\ell \tau_h$, noting that at the level of our study electrons are indistinguishable from muons. We write for each mode a T-odd operator
\begin{equation}
{\cal O}_i=\vec{p}_{\tau^-} \cdot (\vec{p}_{\ell}\times \vec{p}_{j})
\label{semilep}
\end{equation}
and construct a corresponding integrated asymmetry 
\begin{equation}
A_i=c_i \dfrac{r_\tau\ \tilde{r}_\tau}{|r_\tau|^2+|\tilde{r}_\tau|^2}
\label{asemilep}
\end{equation}
where the coefficient $c_i$ is estimated numerically as described above and tabulated in the fourth column. The table shows only leptonic decays on the $\tau^-$ side, but we also checked that the conjugated modes have the same asymmetries. If used on charge specific modes as the ones on the table, the asymmetries are T-odd but not CP odd. True CP odd observables are constructed as in Eq.(\ref{semilep}) where leptons (and corresponding hadronic modes) and anti-leptons are included in the sum. 

\begin{table}
\begin{center}
\begin{tabular}{|c|c|c|c|}
\hline  & Mode & Jets & $c_i$ \\ 
\hline 1 & $(\tau^- \to \nu_\tau \mu^- \bar{\nu}_\mu),(\tau^+ \to \bar{\nu}_\tau \pi^+)$ &  $j=\pi^+$ & -0.27 \\
\hline 2 & $(\tau^- \to \nu_\tau \mu^- \bar{\nu}_\mu),(\tau^+ \to \bar{\nu}_\tau \pi^+ \pi^0)$ & $j=\pi^+ + \pi^0$ & -0.11 \\
\hline 3 & $(\tau^- \to \nu_\tau \mu^- \bar{\nu}_\mu),(\tau^+ \to \bar{\nu}_\tau \pi^+ \pi^0 \pi^0)$ &$j=\pi^+ + \pi^0 + \pi^0$ & -0.017 \\
\hline 4 & $(\tau^- \to \nu_\tau \mu^- \bar{\nu}_\mu),(\tau^+ \to \bar{\nu}_\tau \pi^+ \pi^+ \pi^-)$ &  $j=\pi^+ + \pi^+ + \pi^-$ &  0.0005\\
\hline 
\end{tabular}
\caption{Semi-leptonic modes with tau-jet producing the largest asymmetry and their respective coefficients $c_i$ for Eq.(\ref{asemilep})}
\label{tabl}
\end{center}
\end{table}
The table indicates that the one and two pion modes have the largest asymmetries by far, so that one loses sensitivity by including higher multiplicity modes in the tau-jet. Of course, the higher multiplicity may actually facilitate the experimental reconstruction of the events or the asymmetries so a full study is needed to reach definitive conclusions.

Table~\ref{tabh} shows the modes with two hadronic tau decays $h\to \tau^+\tau^- \to \tau_h \tau_h$ covering 
one, two and three pion modes. As with the semi-leptonic case we studied several possibilities for the definition of the tau-jet, and found the largest asymmetries for the ones shown in the table. We write for each mode a T-odd operator
\begin{equation}
{\cal O}_i=\vec{p}_{\tau^-} \cdot (\vec{p}_{j1}\times \vec{p}_{j2})
\label{twohad}
\end{equation}
and construct a corresponding integrated asymmetry Eq.(\ref{asemilep}).

\begin{table}
\begin{center}
\begin{tabular}{|c|c|c|c|}
\hline  & Mode & Jets & $c_i$ \\ 
\hline 1 & $(\tau^- \to \nu_\tau \pi^-) (\tau^+ \to \bar{\nu}_\tau \pi^+)$  & $j_1=\pi^-$ , $j_2=\pi^+$ & 0.79 \\ 
\hline 2 & $(\tau^- \to \nu_\tau \pi^-),(\tau^+ \to \bar{\nu}_\tau \pi^+ \pi^0)$ & $j_1=\pi^-$ , $j_2=\pi^+ + \pi^0$ & 0.33 \\ 
\hline 3 & $(\tau^- \to \nu_\tau \pi^- \pi^0),(\tau^+ \to \bar{\nu}_\tau \pi^+ \pi^0)$ & $j_1=\pi^- + \pi^0$ , $j_2=\pi^+ + \pi^0$  & 0.13 \\ 
\hline 4 & $(\tau^- \to \nu_\tau \pi^-),(\tau^+ \to \bar{\nu}_\tau \pi^+ \pi^0 \pi^0)$ & $j_1=\pi^-$ , $j_2=\pi^+ + \pi^0 + \pi^0$  & 0.06 \\ 
\hline 5 & $(\tau^- \to \nu_\tau \pi^-),(\tau^+ \to \bar{\nu}_\tau \pi^+ \pi^+ \pi^-)$ & $j_1=\pi^-$ , $j_2=\pi^+ + \pi^+ + \pi^-$  & 0.06 \\ 
\hline 6 & $(\tau^- \to \nu_\tau \pi^- \pi^0),(\tau^+ \to \bar{\nu}_\tau \pi^+ \pi^0 \pi^0)$ & $j_1=\pi^- + \pi^0$ , $j_2=\pi^+ + \pi^0 + \pi^0$ & 0.02 \\ 
\hline 7 & $(\tau^- \to \nu_\tau \pi^- \pi^0),(\tau^+ \to \bar{\nu}_\tau \pi^+ \pi^+ \pi^-)$ & $j_1=\pi^- + \pi^0$ , $j_2=\pi^+ + \pi^+ + \pi^-$ & 0.02 \\
\hline 8 & $(\tau^- \to \nu_\tau \pi^- \pi^0 \pi^0),(\tau^+ \to \bar{\nu}_\tau \pi^+ \pi^0 \pi^0)$ & $j_1=\pi^- + \pi^0 + \pi^0$ , $j_2=\pi^+ + \pi^0 + \pi^0$ & 0.004 \\ 
\hline 9 & $(\tau^- \to \nu_\tau \pi^- \pi^0 \pi^0),(\tau^+ \to \bar{\nu}_\tau \pi^+ \pi^+ \pi^-)$ & $j_1=\pi^- + \pi^0 + \pi^0$ , $j_2=\pi^+ + \pi^+ + \pi^-$ & 0.003 \\ 
\hline 10 & $(\tau^- \to \nu_\tau \pi^- \pi^+ \pi^-),(\tau^+ \to \bar{\nu}_\tau \pi^+ \pi^+ \pi^-)$& $j_1=\pi^- + \pi^+ + \pi^-$ , $j_2=\pi^+ + \pi^+ + \pi^-$ & 0.003 \\
\hline 
\end{tabular}
\caption{Double hadronic tau decays with tau-jets producing the largest asymmetry and their respective coefficients $c_i$ for Eq.(\ref{asemilep}).}
\label{tabh}
\end{center}
\end{table}
As with the semi-leptonic case, we have not listed all the conjugate modes. If the counting asymmetry is constructed for a particular (not self-conjugate) mode, the result is T-odd but not necessarily CP odd. However, if sums over conjugate modes are considered, then any non-zero asymmetry signals CP violation. We find here also that the most sensitive modes are those with only one or two pions.

\section{Conclusions}

In the SM the higgs boson does not have LFV decays and its decays conserve CP. We have argued generically that if one goes BSM to allow LFV decays of the Higgs such as the one suggested by a recent CMS result, one also introduces CP violation. The only channel where it is in principle possible to study this CP violation at LHC is $h\to \tau\tau$ and we have studied the relative sensitivity of different tau-lepton decay modes to CP violating couplings. We have constructed to specific multi-Higgs models in which the 125 GeV Higgs can have LFV decays and argued that only one of them exhibits CP violation as well. These two examples illustrate the different ingredients that are needed for both effects to appear BSM. 

The correlation between LFV and CPV couplings depends on the details of the flavour sector BSM and we have considered two benchmark scenarios. In the first one, the lepton flavor sector has a dominant hierarchical structure that produces the charged lepton masses, but the deviations from this are democratic. We found that in this case the tightest constraint on  possible new physics arises from bounds on $h\to \mu\mu$ and $h\to ee$. Within factors of two, this constraint is consistent with the upper bound on LFV from CMS, and allows for a CP violating asymmetry as large as 11\%. 

In the second benchmark scenario we assumed the corrections to the SM lepton flavor sector are also hierarchical as in the Fritzsch model. In this case the tightest constraints on new physics arise from $h\to \tau\tau$. Within factors of two they are consistent with the upper bound on LFV from CMS, and they allow for a CP violating asymmetry as large as 40\%.

\begin{acknowledgments}

The work of G.V. was supported in part by the DOE under contract number DE-SC0009974. X-G He was supported in part by MOE Academic Excellent Program (Grant No.~102R891505), NCTS and MOST of ROC (Grant No.~MOST104-2112-M-002-015-MY3), and in part by NSFC (Grant Nos.~11175115 and 11575111) and Shanghai Science and Technology Commission (Grant No.~11DZ2260700) of PRC.  X.~G.~H. thanks Korea Institute for Advanced Study (KIAS) for their hospitality and partial support while this work was 
completed. We thank Tao Han for useful information concerning Higgs physics at the ILC.

\end{acknowledgments}

\end{document}